\documentclass[twocolumn,rmp,aps,showpacs]{revtex4}

\usepackage{dcolumn,graphicx,amsmath,amssymb,pxfonts}

\begin{document}

\title{Network dynamics of ongoing social relationships}
\author{Petter \surname{Holme}}
\email{holme@tp.umu.se}
\affiliation{Department of Physics, Ume{\aa} University, 
  901~87 Ume{\aa}, Sweden}

\begin{abstract}
  Many recent large-scale studies of interaction networks have
  focused on networks of accumulated contacts. In this paper we
  explore social networks of ongoing relationships with an emphasis on
  dynamical aspects. We find a distribution of response times (times
  between consecutive contacts of different direction between two
  actors) that has a power-law shape over a large range. We also argue
  that the distribution of relationship duration (the time between the
  first and last contacts between actors) is exponentially
  decaying. Methods to reanalyze the data to compensate for the finite
  sampling time are proposed. We find that the degree distribution for
  networks of ongoing contacts fits better to a power-law than the
  degree distribution of the network of accumulated contacts do. We
  see that the clustering and assortative mixing coefficients are of
  the same order for networks of ongoing and accumulated contacts, and
  that the structural fluctuations of the former are rather large.
\end{abstract}

\maketitle

\section{Introduction}

The recent development in database technology has allowed researchers
to extract very large data sets of human interaction sequences. These
large data sets are suitable to the methods and modeling techniques
of statistical physics, and thus, the last years has witnessed the
appearance of an interdisciplinary field between physics and
sociology~\cite{ba:rev,doromen:rev,mejn:rev}. More specifically these
studies have focused
on network structure---in what ways the networks of
social interaction deviates from completely random networks, and how
this structure can emerge from individual behavior. Most\footnote{To our knowledge, the only type of large (relatively)
  instantaneous network figuring in recent physics literature is
  networks of corporate directors sitting in the same board, Ref.~\cite{davis:board0}.} of these
recent large-scale social network studies have
focused on networks of accumulated relationships. In many cases, the
social network of interest is rather the network of ongoing social
relationships: The dynamics of the spreading of
diseases~\cite{andersonmay}, opinion formation~\cite{bahr:op}, and 
fads~\cite{watts:fad} are often rather fast compared to the evolution
of the network---in such cases
inactive relationships have no relevance. In social search
processes~\cite{watts:search}, distant acquaintances can be helpful,
but not all acquaintances a person has ever had. We also
believe the network of ongoing contacts lies conceptually closer to
the colloquial idea of a network of friends, than what the network of
actors and their accumulated contacts do. Furthermore, traditional social
network studies (e.g.\ Refs.~\cite{wiring,deep,rap:school}) based on
interviews or field surveys has mapped out 
ongoing contacts. The complication, and
probably the reason earlier studies have focused on the network of
accumulated contacts, is that the time of a tie's cessation is less
clear-cut than its beginning. However, if the sampling time of
the data set is very large compared to the network dynamics; then one
can, at a time $t$ in the interior of the sampling time span, approximate the
network of ongoing relationships by the network of contacts that has
occurred and will occur again. In the present paper we use this
method to study the structure and structural fluctuations of
networks of ongoing relationships. To justify that the sampling time
is long enough compared to the time evolution of the network,
we investigate the temporal structure of the relationships. The
data sets we use are obtained from scientific
collaborations~\cite{mejn:scicolpre1}, email
exchange~\cite{bornholdt:email} and interaction within an Internet
community~\cite{pok}.

\section{Notations and network construction}

All our data sets take the form of lists of triples, or
\emph{contacts}, $(v_A,v_B,t)$ meaning that $v_A$ and $v_B$ has
interacted at time $t$. For the scientific collaboration networks the
two first arguments are unordered, for the other two networks the
interaction is directed. We call the set of contacts with the same two
first elements (neglecting the order) a \emph{relationship} between
$v_A$ and $v_B$. Our approximation of the graph of ongoing contacts at
time $t$ is then defined as $G(t)=\{V(t),E(t)\}$, where $V(t)$ is the
set of vertices (or actors) that occur in a contact at a time earlier
than $t$, and $E(t)$ is the set of unordered pairs of vertices
$(v_A,v_B)$ where there exist contacts between $v_A$ and $v_B$ at
times $t'$ and $t''$ such that $t'<t<t''$.

\begin{table*}
\caption{Statistics of the networks. Date notations have the format
  year-month-day hour:minute:second GMT. The number of ties does not
  include self-communication (e.g.\ self-addressed e-mails) but
  in ``number of contacts'' such communication is included.}
\label{tab:stat}
\begin{ruledtabular}
\begin{tabular}{r|lll}
   & e-prints & e-mail & pussokram.com \\\hline
start of sampling  & 1995-01-01 06:00:00 & 2001-07-29 03:11:33 &
2001-02-13 14:39:25\\ 
end of sampling & 2001-01-01 05:31:00 & 2001-11-18 02:06:28 &
2002-07-10 15:28:00\\
sampling duration, $t_\mathrm{stop}$ & $2192.0$ days & $112.0$ days&
$512.0$ days\\
number of actors, $N$ & $58{\,}342$& $64{\,}370$ & $29{\,}341$\\
number of ties, $M$ & $294{\,}901$ & $97{\,}425$ & $115{\,}684$\\
number of contacts & $530{\,}481$ & $447{\,}543$ & $536{\,}276$\\
relationship duration, $t_\mathrm{dur}$  & 1532(30) days & 187(5)
days & 129(10) days
\end{tabular}
\end{ruledtabular}
\end{table*}

For the network of scientific collaborations we use similar data
as used in Ref.~\cite{mejn:scicolpre1} (but sampled one year
longer). This data is extracted from the preprint repository arxiv.org
where scientists themselves can upload manuscripts. An edge between
$v_A$ and $v_B$ means that $v_A$ has appeared as a coauthor of a
preprint together with $v_B$. The time the manuscript is uploaded is
the time we say the collaboration has occurred.

The email network is the same data set as presented in
Ref.~\cite{bornholdt:email} and consists of all in- and out-going
email traffic to a server handling undergraduate students'
email accounts in Kiel, Germany.

The Internet community network is constructed from the same data set
as in Ref.~\cite{pok}. Here an edge represents any of four different
ways of contacts between users of the Swedish Internet community
pussokram.com. This community is intended for romantic communication
among adolescents and young adults.

For the email and pussokram.com networks one can define a direction
for the contacts. In the study of network structure, however, we will
consider the contacts as bidirectional. Statistics for the networks
are presented in Table~\ref{tab:stat}.

\begin{figure}
  \centering{\resizebox*{8cm}{!}{\includegraphics{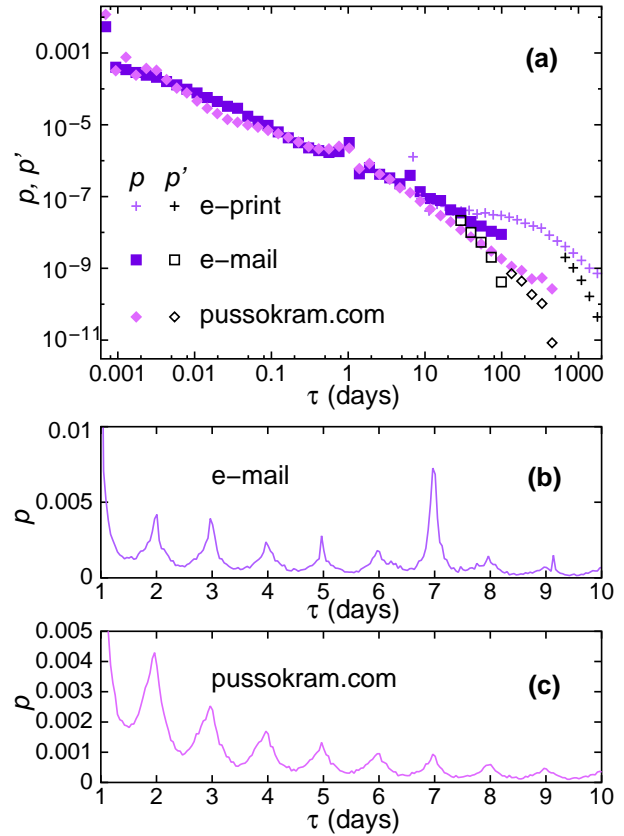}}}
\caption{Response time statistics. $p'$ is the frequency of
  response times of in the data sets. $p$ is the recalculated quantity
  compensated for the finite sampling time. (a) shows $p$ for
  all data sets in log scales. The data is log-binned. $p'$ is plotted
  for the last five bins, elsewhere $p'$ and $p$ overlaps to a great
  extent. A blow-up of $p$ (in linear scale) is shown in (b) (e-mails)
  and (c) (pussokram.com).}
\label{fig:resp}
\end{figure}

\section{Relationship dynamics and the speed of network change}

Before we investigate the approximate network of ongoing contacts, as
defined above, we discuss the speed of interaction and the validity of
the approximation. First, we focus on the distribution of response
times $\tau$---times between consecutive contacts of different
direction within a relationship.\footnote{As mentioned we will focus
  on undirected networks later, but for comparison with other works we
use directed contacts in this definition. The conclusion from an
undirected definition would be the same.} For the undirected e-print
data we
simply define $\tau$ as the time between consecutive uploads of
e-prints within a relationship. We measure the
$\tau$-distribution of the data sets, $p'$, and also a
quantity $p$ where the effects of the finite size effects are
compensated for. An earlier study~\cite{eckmann:dialog} has found a
power-law like $\tau$-distribution. As shown in Fig.~\ref{fig:resp}(a)
this picture is confirmed in the large scale. This stretched
functional form makes the finite sampling time a problem as it
imposes a cut-off on the recorded distribution $p'$. To compensate
for this and construct a better approximation $p$ to the real
distribution, we use the formula $P(B_\tau) = P(A\cap B_\tau)/ P(A
|B_\tau)$ where $A$ is the event that a response interval that starts
within the sampling interval $I_t=[0,t_\mathrm{stop}]$
also ends within $I_t$, and $B_\tau$ is the statement that the
response time is $\tau$. Now $P(A\cap B_\tau)$ is just the frequency
distribution of interval length as measured during the sampling. To
find $P(A|B_\tau)$ we note that, if we assume that contacts occurs
with a constant rate (which is reasonable in a long term perspective
for a system of a relatively constant number of actors), then a
response interval ends within $I_t$ with probability
\begin{equation}
1-\tau/t_\mathrm{stop} ~ .\label{eq:pr1}
\end{equation}
However, the sizes of the communities need not to be time independent.
For response intervals involving an actor that enters the system at
time $t$ Eq.~(\ref{eq:pr1}) becomes $1-(\tau+t)/t_\mathrm{stop}$. Now
we approximate the time an actor $v$ enters the system with the first
time $t_v$ that $v$ is involved in a contact, and get the formula:
\begin{equation}
  p(\tau)=P(A|B_\tau ) = a \sum_{v\in V} \Theta
  (t_\mathrm{stop}-t_v-\tau)
  \left[1-\frac{t_v+\tau}{t_\mathrm{stop}}\right]
  \label{eq:prcond}
\end{equation}
where $\Theta(\; \cdot\; )$ denotes the Heaviside function, and $a$ is
a normalizing constant. $p$ is plotted, along with the
$p'$ values that differs most from $p$, in Fig.~\ref{fig:resp}(a)
(here $a$ is chosen to make $p$ coincide with $p'$ for small $\tau$
values rather than to normalize
$p$). We note that $p$ is straighter than $p'$ in the log-log scale
(for at least the e-mail and pussokram.com curves),
which suggests a power-law like behavior over a considerable
range. (Of course there is eventually a cut-off---from the human life
time, if nothing else.) The e-print curve has a peculiar bend as it
seems to shift exponent around $\tau=300\:\mathrm{days}$, an
observation we hope future studies can explain. There is a conspicuous
irregularity around $\tau=1\:\mathrm{day}$ for the e-mail and
pussokram.com curves. This was also observed in Ref.~\cite{eckmann:dialog} and
explained as an effect of people's everyday routines---the Kiel
students read and reply their e-mails at the same hour as the day before,
the pussokram.com members log in after school or work, and so on. This
effect is more visible in a linear scale, see Figs.~\ref{fig:resp}(b)
and (c). For the e-mail curve the peak at seven days is larger than
the surrounding peaks, indicating that some emails are associated with
weekly routines among the Kiel students and their contacts. This
one-week-peak can not be seen in the pussokram.com curve; possibly
reflecting that business (or university studies) has more weekly
scheduled routines than leisure do. 

\begin{figure*}
  \centering{\resizebox*{15cm}{!}{\includegraphics{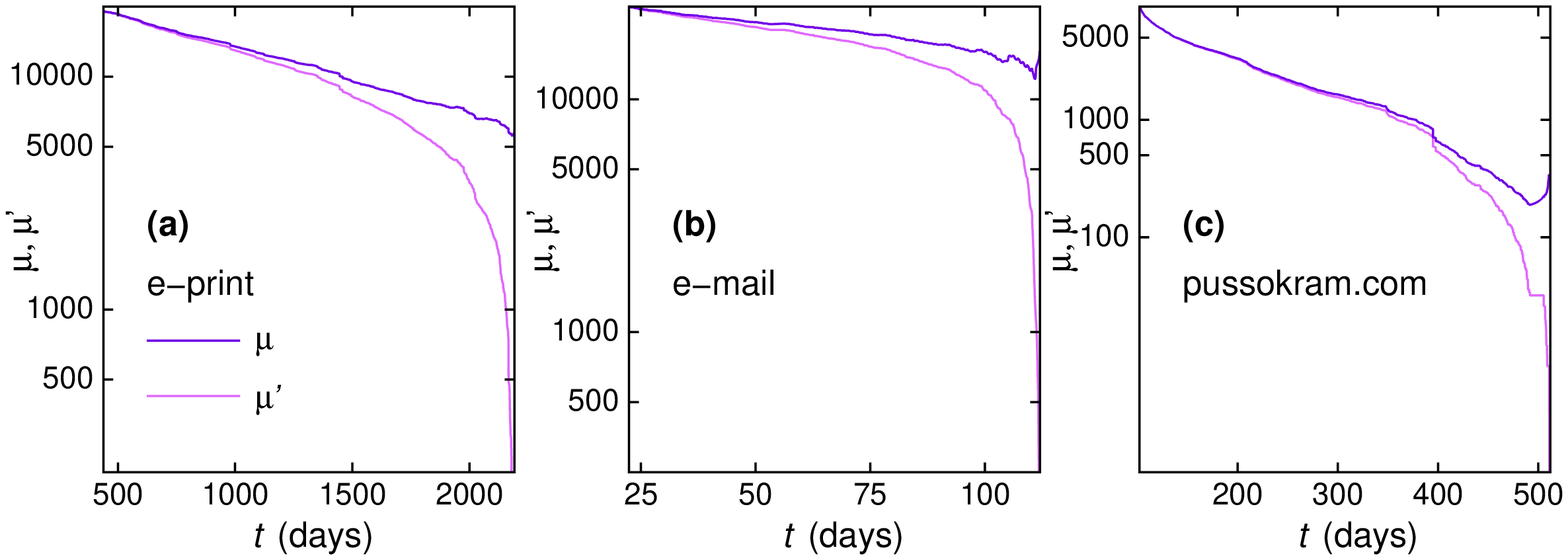}}}
\caption{The speed of network change. $\mu'$ is the number of edges at
  time $t_0$ that still are present at time $t>t_0$. We choose
  $t_0=0.2 t_\mathrm{stop}$. }
\label{fig:m}
\end{figure*}

Now we turn to the more central question about the speed of
relationship cessation. Our central quantity is the number of
relationships existing at time $t_0$ that still remains at time $t$
(we assume $0\leq t_0 <
t < t_\mathrm{stop}$), $\mu(t_0,t)$. This quantity can crudely be
approximated with the number of relationships at $t$ that existed at
$t_0$ that will occur again before $t_\mathrm{stop}$,
$\mu'(t_0,t)$. The error in the approximation will be rather large for
$t$ close to $t_\mathrm{stop}$. But, just as above, one can improve
the approximation considerably. If one assumes that the response time
distribution $p(\tau)$ applies to all relationships regardless if the
relationship is new or old; then, during a time interval $\Delta t$,
the change of $\mu$ can be written:
\begin{equation}
  \Delta \mu=\Delta \mu'+ \mu\,\Delta\pi \label{eq:himp}
\end{equation}
where $\pi(t)$ is the probability that a relationship, that has its
last recorded contact at time $t$, actually continues after
$t_\mathrm{stop}$:
\begin{equation}
  \pi(t)=\sum_{\tau=t_\mathrm{stop}-t}^\infty p(\tau)\,\Delta \tau~,
  \label{eq:diffmu}
\end{equation}
where the sum is over the bins of the $p(\tau)$ histogram. A change of
variables gives:
\begin{equation}
  \Delta \pi(t)=\Delta t \, p(t_\mathrm{stop}-t)~,
  \label{eq:ppi}
\end{equation}
and finally a formula for integrating $\mu$:
\begin{equation}
  \mu(t+\Delta t)=\frac{\Delta \mu'(t+\Delta t)+\mu(t)}{1-
  \Delta t \, p(t_\mathrm{stop}-t)} ~.
  \label{eq:mu}
\end{equation}
 We also need the factor $a$ of
Eq.~(\ref{eq:prcond}) which is hard to estimate since we don't exactly know
$p(\tau)$'s long term behavior. However, we note that for certain $a$
the $\mu(t)$ curves are rather straight in a lin-log plot, see
Fig.~\ref{fig:m} (as opposed to the e-mail and pussokram.com curves
the e-print curve decays so slowly that a power-law form of $\mu(t)$
cannot be ruled out). This means that the characteristic duration time
is well-defined---fitting to an exponential
$A\exp(-t/t_\mathrm{dur})$  ($A$ and $t_\mathrm{dur}$ are the two
degrees of freedom) gives the characteristic durations
$t_\mathrm{dur}$ of relationships displayed in
Table~\ref{tab:stat}. To be able to approximate the network of ongoing
contacts with the network of contacts that have happened and will
happen again one would like $t_\mathrm{dur} \ll t_\mathrm{stop}$ to
hold. We see that for the pussokram.com data $t_\mathrm{stop}$ is
about four times as large as $t_\mathrm{dur}$ which enables us to draw
some conclusions using this approximation. The effective sampling
times of the e-print and e-mail data are, however, so short that we
exclude these for the latter section of this paper.

\begin{figure}
  \centering{\resizebox*{8cm}{!}{\includegraphics{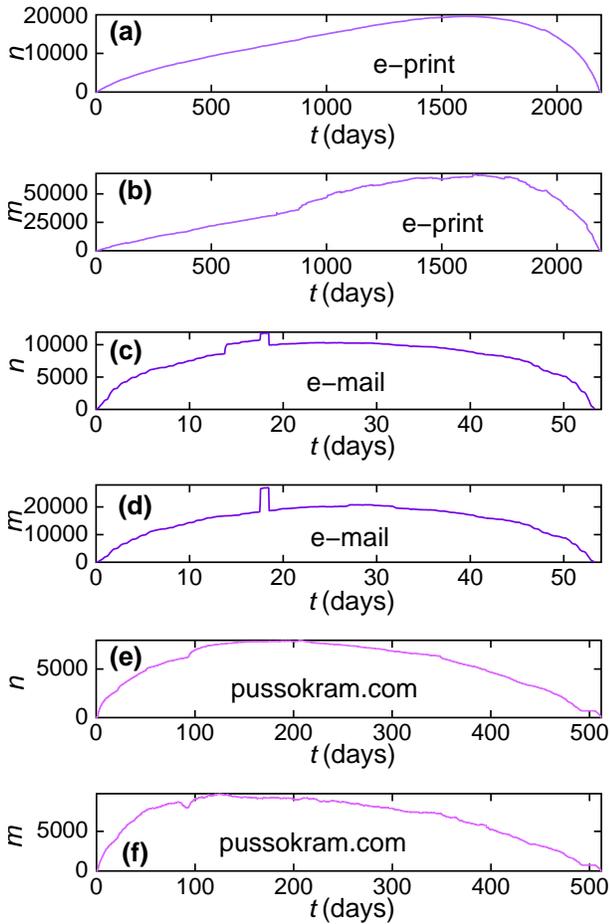}}}
\caption{(a) shows the number of vertices $n(t)$ and (b) shows the
  number of edges $m(t)$ of the networks of contacts that have occurred
  and will occur again.}
\label{fig:size}
\end{figure}

Now we take a brief look at the time evolution of the network
sizes---$n$, the number of active actors at time $t$, and $m$, the
number of edges in our approximate network of ongoing relationships. If the
number of active users increases during the sampling period, the time
evolution of $n$ and $m$ should be right-skewed, and this is indeed
true for the e-print data (as seen in Figs.~\ref{fig:size}(a) and
(b)). The e-mail and pussokram.com curves are more symmetric (the
pussokram.com curve is indeed slightly left-skewed). We note that
for pussokram.com, $m$ is much less than $M$. The kinks of the e-mail
curves are due to group or spam e-mails, the other quantities $m$, $p$
and so on, are not affected by this.

\begin{figure*}
  \centering{\resizebox*{15cm}{!}{\includegraphics{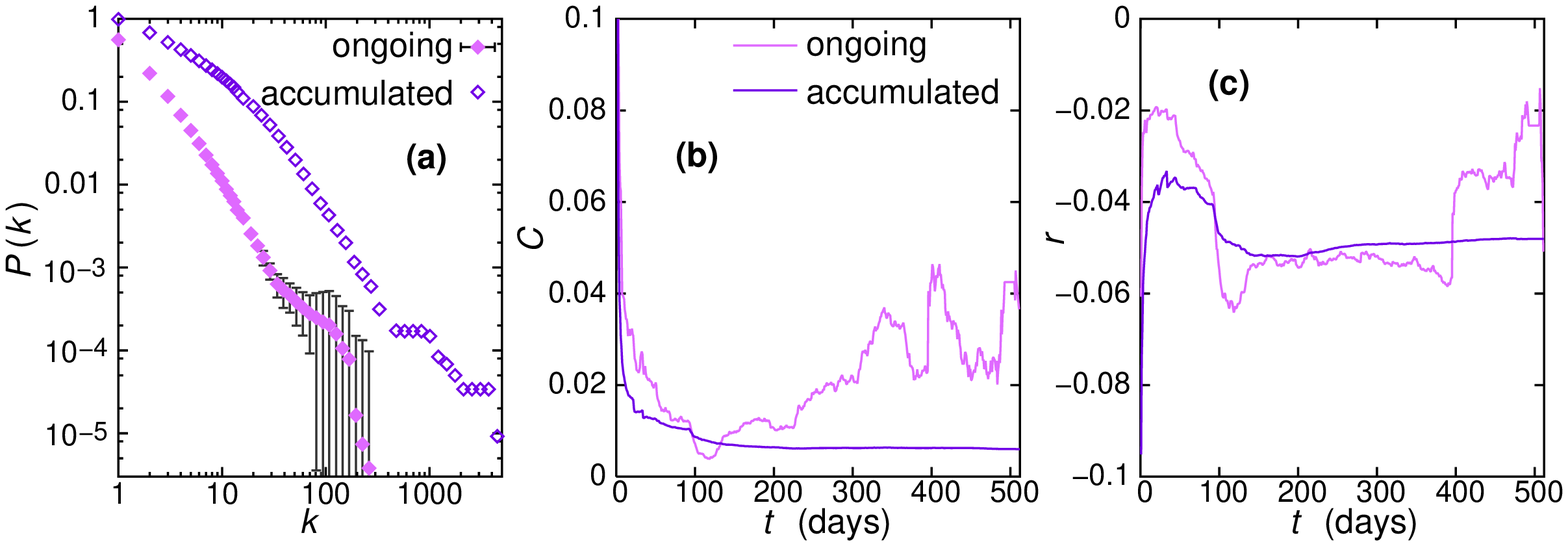}}}
\caption{Structure of the network of ongoing contacts. (a) shows the
  cumulative degree distribution of the network of ongoing contacts along with
  the degree distribution of the network of accumulated
  contacts. Error-bars are shown if larger than symbol size. The errors
  are calculated assuming that networks at times differing more than
  $t_\mathrm{dur}$ are independent. (b) shows the clustering
  coefficient as function of time for networks of ongoing and
  accumulated contacts. (c) shows assortative mixing coefficient as
  function of time.}
\label{fig:s}
\end{figure*}

\section{Network structure and structural fluctuations}

Now we turn to the structure of the network of ongoing contacts, and
the fluctuations of the structural measures. In this Section we only
use the pussokram.com data (due to, as mentioned above, the large
effective sampling time for this data set). We focus on three
quantities that recently have received much attention:  The first
structural measure is the distribution of degree (number of edges to a
vertex). The first quantity is the clustering coefficient $C(G)$ where
we use the traditional sociological definition
\begin{equation}
  C(G)=c_3(G)/p_3(G)~,
\end{equation}
where $c_3(G)$ is the number of representations of every 3-cycle
(triangle) of $G$\footnote{If $(v_1,v_2,v_3)$ is a
  triangle, then $(v_2,v_3,v_1)$ another representation of the same
  triangle. So the number of distinct  triangles is $c_3(G)/6$.} and
$p_3(G)$ is the number of representations of 3-paths. The second
quantity is the assortative mixing coefficient~\cite{mejn:assmix}
\begin{equation}
  r=\frac{\langle k_1\, k_2\rangle -
    \langle k_1\rangle  \langle k_2\rangle}
  {\sqrt{\langle k_1^2\rangle - \langle
    k_1\rangle^2}
    \sqrt{\langle k_2^2\rangle- \langle
      k_2\rangle^2}}
\end{equation}
where averages are taken over $E$, and $k_1$ and $k_2$ are the degrees
of an edge's first and second arguments as they appear in $E$.

The cumulative degree distribution of our approximate network of ongoing
relationships, along with the corresponding data for the network of
accumulated contacts is plotted in Fig.~\ref{fig:s}(a). Just as for
the accumulated network, our approximate network of ongoing
relationships has a fat tailed degree distribution; but the network of
ongoing relationships fits better to a single power-law with. The
stronger downward bend of $P(k)$ for accumulated social contacts has
been observed earlier~\cite{mejn:scicolpre1,pok}; maybe this larger
correction to a power-law form is due to inactive edges. We note that
even if the degree distribution fits very well to that of the
Barab\'{a}si-Albert model~\cite{ba:model},\footnote{For a case study
  of papers citing Ref.~\cite{ba:model}, 
  see Ref~\cite{putsch}.} the central
ingredient in the Barab\'{a}si-Albert model (the ``preferential
attachment'') does not apply directly to the pussokram.com
community. Preferential attachment means that a vertex acquires new
edges with a rate proportional to its degree, but in the pussokram.com
community the degree of a member is invisible to others~\cite{pok}.

Next we turn to the time evolution of $C$ and $r$. In
Figs.~\ref{fig:s}(b) and (c) these are displayed for the whole
sampling time. The earliest and latest times can, of course, be
affected by the proximity to the borders of the sampling time
frame---for our discussion we focus on the interval
$[t_\mathrm{dur}, t_\mathrm{stop}-t_\mathrm{dur}]\approx
[129,383]\:\mathrm{days}$. We see that both $C$ and $r$ are of the
same order of magnitude for the networks of ongoing and accumulated
contacts. These values of $C$ and $r$ are rather neutral in the sense
that they can be expected from a random network with a skewed degree
distribution~\cite{mejn:why}. The fluctuations are rather large,
especially for the clustering coefficient (with a standard deviation
of around half the average value). An intriguing question for
future studies is how dynamical systems on the networks are affected
by strong structural fluctuations. Slightly outside our interval, at
$t\approx 395$ there is an upward jump in both $C$ (from $0.023$ to
$0.041$) and $r$ (from $-0.057$ to $-0.043$) that is the result of a
new contact between two of the most central actors. Such a new edge introduces
a new triangle for every common neighbor of the two vertices, and
can thus increase $C$ substantially as two high-degree actors may have
many neighbors in common. Such an event will, by definition, also give
a large positive contribution to the assortative mixing.
We can expect sudden jumps in many structural quantities for networks
of ongoing relationships with fat tailed degree distributions, as the
rare event of an edge appearing or disappearing between two of the
most connected vertices will affect many structural measures (various
kinds of centrality measures~\cite{wf} are probably even more sensitive
to such events). Unfortunately the sampling time is too short,
despite the fast pussokram.com dynamics, to get good statistics for
the autocorrelation function of $C(t)$ and $k(t)$ (it is 
consistent with a characteristic time of decay similar to
$t_\mathrm{dur}$).

\section{Summary and Conclusions}

In this paper we investigate networks of ongoing contacts from three
large sets of social interaction data. We study the response time
distribution and distribution of relationship duration. We reanalyze
these quantities to compensate for the finite sampling time by
supposing that the response time distribution is the same for all
relationships, and the same throughout the duration of the
relationship. We find a response time distribution that has a
power-law like shape in the large scale, but has an informative
small-scale structure reflecting the daily and weekly routines. The
distribution of relationship duration is consistent with an
exponential decay. This indicates that there is a well-defined
characteristic duration time of a relationship, $t_\mathrm{dur}$; and
that if $t_\mathrm{dur}$ is much less than the sampling time
$t_\mathrm{stop}$ the network of ongoing contacts can be reasonably
well approximated by the network of contacts that have happened and
will happen again. For one of our data sets---that of the Internet
community pussokram.com---we have $4 t_\mathrm{dur}\approx
t_\mathrm{stop}$. For this data set we compare the approximate network
of ongoing contacts with networks of accumulated contacts---the common
way of constructing social networks from interaction data. We find a
degree distribution that fits much better to a power-law for the
network of ongoing contacts than the network of accumulated
contacts. The clustering coefficient and assortative mixing
coefficients are of the same order; which, to some extent, justifies
the use of network of accumulated contacts as a proxy for networks of
ongoing contacts. The fluctuations in these quantities are, however,
rather large. A fact that may have important consequences for
dynamical systems. We hope these results will inspire more extensive
longitudinal studies of interaction networks with fast dynamics, as
well-converged data for relationship duration distribution and
autocorrelation functions of structural quantities are within
reach. We also point out the interplay between dynamical systems on
the networks and the structural fluctuations as an interesting area of
future studies.

\section*{Acknowledgments}

The author would like to thank Niklas Angemyr,
Stefan Bornholdt, Holger Ebel, Michael Lokner,
Mark Newman, and Christian Wollter for help with data
acquisition; and Beom Jun Kim for comments and suggestions. The
author was partly supported by the Swedish Research Council through
contract no.\ 2002-4135.

\end{document}